\documentclass[showpacs,twocolumn,aps,amsfonts]{revtex4}

\usepackage{mathrsfs}
\usepackage{epsfig}
\usepackage{psfig}
\usepackage{bm}

\newcommand{\be}{\begin{equation}}
\newcommand{\ee}{\end{equation}}
\newcommand{\bea}{\begin{eqnarray}}
\newcommand{\eea}{\end{eqnarray}}

\def\la {\langle}
\def\ra {\rangle}
\def\curl{\mathsf{curl\,}}
\newcommand{\D}{{\mathrm{D}}}

\newcommand{\bra}[1]{\left(#1\right)}
\newcommand{\bras}[1]{\left[#1\right]}

\def\a{\alpha}
\def\b{\beta}
\def\g{\gamma}
\def\d{\delta}
\def\s{\sigma}
\def\e{\epsilon}
\newcommand{\sfr}[2]{{\textstyle\frac{#1}{#2}}}
\newcommand{\reff}[1]{(\ref{#1})}

\begin{document}

\title{Perfect magnetohydrodynamics as a field theory}
\author{Jacob D. Bekenstein and Gerold Betschart}
\affiliation{Racah Institute of Physics, Hebrew University of
Jerusalem, Jerusalem 91904, Israel}

\begin{abstract}

We propose the generally covariant action for the theory  of a self-coupled complex scalar field and electromagnetism which by virtue of constraints is equivalent, in the regime of long wavelengths, to perfect magnetohydrodynamics (MHD).   We recover from it the Euler equation with Lorentz force, and the thermodynamic relations for a prefect fluid.  The equation of state of the latter is related to the scalar field's  self potential.  We introduce $1+3$ notation to elucidate the relation between MHD and field variables.  In our approach the requirement that the scalar field be single valued leads to the quantization of a certain circulation in steps of $\hbar$; this feature leads, in the classical limit,  to the conservation of that circulation.  The circulation is identical to that in  Oron's generalization of Kelvin's circulation theorem to perfect MHD; we here characterize the new conserved helicity associated with it. We also demonstrate the existence for MHD  of two Bernoulli-like theorems for each spacetime symmetry of the flow and geometry; one of these is pertinent to suitably defined potential flow.  We exhibit the conserved quantities explicitly in the case that two symmetries are simultaneously present, and give examples.  Also in this case we exhibit a new conserved MHD circulation distinct from Oron's, and provide an example.

\end{abstract}



\date{\today}

\maketitle

\section{Introduction}\label{intro}

It is well known that perfect fluid dynamics admits a description in terms of dynamics of a self-interacting real scalar field.   Analogously, a superconductor, insofar as we are interested in the flow of current through it, can be viewed as a complex (charged) scalar field with self-interaction coupled to the electromagnetic field~\cite{LG}.  Can the flow of a plasma interacting with the electromagnetic field be treated analogously ?  In one approach such a plasma is represented by two charged Schr\" odinger-like scalar fields representing the flows of ions and electrons, respectively~\cite{faddeev}.

Here we ask, is perfect magnetohydrodynamic (henceforth MHD) flow, namely flow of plasma with negligible charge separation, and with a magnetic field  frozen into the flow, amenable to relativistic description in terms of the dynamics of a single, possibly complex, scalar field interacting with the electromagnetic field ?  Such a description would not only be of methodological interest, but might supply new insights into
MHD flow as well as hints about solution of its intricate equations in a wider range of problems than possible hitherto.

In this paper we approach the problem by writing a joint action for the electromagnetic field and a complex scalar field, the latter self-coupled as well as coupled to a non gauge vector potential.   Part of the action, which is generally covariant,  involves constraints, one of which enforces  the condition for the magnetic field to be frozen into the flow.   We show that in the long wavelength limit the dynamics following from this action is that of MHD. Among the immediate consequences of the field point of view are concise derivations of the circulation conservation theorem in MHD, and of Bernoulli-like theorems, one for each spacetime symmetry.

Our approach offers a subtle resolution to a quandary originating in scalar field formulations of hydrodynamics.  Let us understand the problem.   The energy-momentum tensor for a relativistic real scalar field whose gradient is timelike can be cast  in the form of the energy-momentum tensor for a fluid whose density and pressure are related to the self-interaction potential, and whose 4-velocity  is the gradient of a function of the scalar field.   Thus only potential flow is so described.  It would seem that a scalar field representation of hydrodynamics is incapable of covering all types of flow, even if we grant the perfect flow condition.

It is interesting to contrast this drawback with a similar one evidenced by the purely fluid Lagrangian formulation of  isentropic hydrodynamic flow~\cite{herivel,eckart,mittag,taub,kodama}.  The velocity field is there the sum of two terms: the gradient of the Lagrange multiplier responsible for enforcing mass conservation, and the entropy per unit mass times the gradient of the Lagrange multiplier responsible for enforcing entropy conservation (which is a prerequisite for isentropic flow).   The velocity field is thus again potential.

The above problem is usually fixed by invoking Lin's trick~\cite{lin}: introduce a new Lagrange constraint enforcing the conservation of some additional local quantity along the flow.  In this case the velocity field is generic and not restricted to potential flow~\cite{seliger,schutz,BAO}.  However, the nature of the extra conservation law is mysterious.  It is usually claimed that the conserved quantity is one of the Lagrangian coordinates of fluid elements.  But then the question arises, why only one of the three such coordinates enters.  And even if there were a principle that chose one of the three for its special role, one may ask, why cannot the problem be formulated entirely in Eulerian coordinates ?

In the approach proposed here the fluid 4-velocity is a scalar quantity times the difference of the scalar field's gradient and the 4-potential (not a gauge one).  The 4-velocity is not collinear with any gradient because the phase, being a modular variable, does not necessarily yield a true gradient (one whose curl vanishes) upon differentiation.  This puts both vortical and potential flows within the province of the theory.

In Sec.~\ref{content} we define the scalar and electrodynamic variables of the theory, and propose the action including the constraints which establish the connection with MHD flow.  In Sec.~\ref{eqns} we derive from the least action principle the theory's equations, including the source of Maxwell's equations and the MHD Euler equation of motion.  We also discuss  the gauge freedom of the theory, showing that it is limited to a $U(1)\times U(1)$ group.   In Sec.~\ref{character} we obtain the equation of state of the fluid represented by the scalar field and its fluid-like energy-momentum tensor. Sec.~\ref{3+1}  provides a translation of the theory into $1+3$ language in order to clarify the theory's content.  For example,  we show how to represent the kinematical variables of the flow in terms of the scalar's phase.  Sec.~\ref{hel} is devoted to Oron's generalization of Kelvin's circulation theorem to perfect MHD~\cite{BEO,BAO} in the scalar field language; it also characterizes the new conserved helicity associated with it.  Finally, in Sec.~\ref{sec:bernoulli} we demonstrate the existence in MHD of two Bernoulli-like theorems for each spacetime symmetry of the flow, one of these applicable to suitably defined potential flow.  We exhibit the conserved quantities  explicitly  in the case that two symmetries are simultaneously present,  and provide examples.  Sec.~\ref{sec:bernoulli} also exhibits a new conserved circulation distinct from Oron's and Kelvin's, and provides an example.

We work in units with $c=1$.  Our signature is $\{-,+,+,+\}$.
Symmetrization and
anti-symmetrization of tensors is denoted by $T_{(\a\b)} \equiv
\sfr12\bra{T_{\a\b}+T_{\b\a}}$ and $T_{[\a\b]} \equiv
\sfr12\bra{T_{\a\b}-T_{\b\a}}$, respectively.  In like manner  $W_{[\a\b;\g]}\equiv \sfr{1}{6}
\{W_{\a\b;\g}+W_{\g\a;\b}+W_{\b\g;\a}-(\a\Longleftrightarrow\b)\}$.

\section{The theory's content}
\label{content}

\subsection{Fluid}\label{fluid}

The four velocity of the MHD fluid is written $u^\alpha$ while its baryon
proper density will be denoted $n$.  Other thermodynamic quantities are the energy proper density $\rho$, the (isotropic) pressure $p$ and the relativistic enthalpy per baryon $\mu$,
\be
\mu={\rho+p\over n}. \label{mu}
\ee

\subsection{Electromagnetic field}\label{EM}

Relatively to $u^\a$, the Faraday tensor $F_{\a\b}$ and its dual
$^*\!F_{\a\b}$ can always be decomposed into electric and
magnetic parts,
\bea
F_{\a\b} &=& ~u_\a E_\b - u_\b E_\a + \varepsilon_{\a\b\g\d}\,u^\g B^\d,
\label{F}
\\
^*\! {F}_{\a\b} &=& -u_\a B_\b + u_\b B_\a + \varepsilon_{\a\b\g\d}\,u^\g
E^\d,
\label{Fstar}
\eea

where the electric field is defined as $E_\a \equiv F_{\a\b}\,u^\b$
and the magnetic field is $B_\a \equiv \sfr12 \varepsilon_{\a\b\g\d}\,u^\g
F^{\b\d} \equiv\,  ^*\! {F}_{\g\a}\,u^\g$, respectively. The Levi-Civita tensor is normalized such that
$\varepsilon_{0123}=\sqrt{-g}$.

In perfect MHD the conductivity of the fluid is assumed so high that it completely suppresses the electric field: $E_\alpha=0$.  Thus in MHD the Faraday tensor is characterized by the condition
\be
F_{\alpha\beta}\,u^\beta=0. \label{MHD}
\ee

For performing variations the Faraday tensor must be expressed in terms of potentials.  Our notation is  $F_{\alpha\beta}=A_{\beta;\alpha}-A_{\alpha;\beta}=A_{\beta,\alpha}-A_{\alpha,\beta}$.

\subsection{Scalar field}\label{scalar}

We shall represent the fluid by a complex scalar field $\psi$ whose basic action is taken to be
\be
S_\psi = -{\scriptstyle 1\over \scriptstyle
2}\int[(\psi,^\alpha-\imath \eta^\alpha)(\psi_{,\alpha}-\imath
\eta_\alpha)^*+V(\psi\psi^*)]\, (-g)^{1/2}\, d^4x,
\label{actionpsi}
\ee
where $\eta_\alpha$ is an auxiliary vector field.  We have added a self-interaction potential $V(\psi\,\psi^*)$ in accordance with experience in representing fluids by scalar fields. Why is this an appropriate action ?   Were $\psi$ here an electrically charged field (charge $e$), we would write its coupling to the gauge potential $A_\alpha$ as above, but with $\eta_\alpha\mapsto e A_\alpha$.  An MHD fluid is a mixture of two oppositely charged fluids, e.g. electrons and ions, so each component would deserve an action like the above, but with opposite signs of $e$.  Since MHD assumes that the net charge in each small volume vanishes, we must contemplate a (perhaps) imperfect cancellation of the gauge interaction.  Thus if we represent the MHD fluid by a single complex scalar field $\psi$, this last must no longer be minimally coupled to $A_\alpha$.  However, we know physically that $\psi$ must have a residual electromagnetic interaction.  The most economic way to write its effective action is as in Eq.~\reff{actionpsi}, where $\eta_\alpha$ is expected to bear some complicated relation to the electromagnetic field [to be deduced in Eq.~\reff{eta} below].  To put $S_\psi$ in final form we write $\psi=\sigma\, e^{\imath \varphi}$ ($\sigma$ and $\varphi$ real).  In addition to the elements already mentioned, we shall have need of an additional vector field $b^\alpha$ to play the role of a Lagrange multiplier charged with the task of enforcing the MHD condition \reff{MHD}.

\subsection{Action}\label{action}

Omitting gravity's dynamics, the action of the theory  is
$S=S_M+S_\psi+S_c+S_{nor}+S_\zeta$ where
\bea
S_M &=&-{1\over 16\pi}\int F^{\alpha\beta}\,F_{\alpha\beta}\, (-g)^{1/2}\, d^4x,
\\
S_\psi &=& -{\scriptstyle 1\over \scriptstyle
2}\int[\sigma,^\alpha\,
\sigma_{,\alpha}+\sigma^2\,(\varphi_,^\alpha-\eta^\alpha)(\varphi_{,\alpha}-\eta_\alpha)
\label{S_phi}
\\
&& \quad \qquad+V(\sigma^2)] (-g)^{1/2}\, d^4x, \nonumber
\\
S_c&=&{ 1\over   4\pi} \int (F_{\alpha\beta} +\kappa\,
g_{\alpha\beta})\, u^\alpha\, b^\beta\,(-g)^{1/2}\, d^4x,
\\
S_{nor}&=&- {\scriptstyle1\over \scriptstyle2}\int \xi\,
n\,(u^\alpha\,u_\alpha+1)\,(-g)^{1/2}\, d^4x,
\\
S_\zeta&=&-\zeta\int n\, u^\alpha\, \eta_\alpha\,(-g)^{1/2}\, d^4x.
\label{actionintegral}
\eea

$S_M$ is the usual Maxwell action; it is to be viewed as functional of
the electromagnetic 4-potential $A_\alpha$.  In accordance with our earlier remarks, the scalar field action $S_\psi$, now rewritten in terms of $\sigma$ and $\varphi$, represents the  MHD fluid.       We do not include a kinetic
term for $\eta_\alpha$ principally because, as we shall see, the apparent gauge
invariance of the theory is a mirage.

The constraint part of the action $S_c$ takes care of the MHD
condition  \reff{MHD}.  Indeed, interpreting
$b^\alpha$ as local Lagrange multiplier, we see this condition
emerges from variation of the first term in $S_c$ with respect to $b^\alpha$.
But this is not correct in itself.  By antisymmetry of
$F_{\alpha\beta}$, $F_{\alpha\beta}\,u^\beta\,u^\a=0$,
i.e. the vector $F_{\alpha\beta}\,u^\beta$ has only three
independent components.  Hence the MHD condition should be derivable
with help of a triplet of Lagrange multipliers; a 4-vector of
them is too much.  Accordingly we subject $b^\alpha$ to the
constraint $g_{\alpha\beta}\, b^\alpha\, u^\beta=0$ by way of
inclusion in $S_c$ of the term with local Lagrange multiplier
$\kappa$.  Thus only three of $b^\alpha$'s components are independent.

The 4-velocity of the fluid $u^\alpha$ is to be determined  by the
theory in terms of scalar field variables. We have no guarantee that
it will be properly normalized. so we impose such normalization as a
Lagrange constraint by the part of the action $S_{nor}$.  The $\xi$
is a local Lagrange multiplier and the factor $n$ is included so
that 4-velocity normalization is not required in regions not
occupied by the fluid.

Addition of $S_\zeta$ to the action is found to be necessary for the correct relation between $u^\alpha$ and the field variables [see Eq.~\reff{velocity}]  to arise; we have found no illuminating intuition for it.  In $S_\zeta$ the $\zeta$ is a constant with
dimensions of action; it will emerge in Sec.~\ref{con-cir} that $\zeta$ can be identified with the quantum of action $\hbar$.

\section{Equations}\label{eqns}

Variations will be carried out with respect to the independent fields:
$A_\alpha, \eta_\alpha, \varphi, \sigma, b^\alpha, u^\alpha,
g_{\alpha\beta}$ and $n$.  We stress that for this purpose
$u^\alpha, b^\alpha, A_\alpha$ and $\eta_\alpha$ are to be varied
independently of the other variables, but e.g. variation of
$u_\alpha$ entails a variation of $u^\alpha$ and of
$g_{\alpha\beta}$.

\subsection{Constraints and conditions}\label{constraints}

Variation of $S$ with respect to $\xi$ gives  $n\,(u^\alpha\,u_\alpha+1)=0$;
thus
\be
u^\alpha\,u_\alpha=-1 \quad {\rm where}\quad n\neq 0
.
\label{norm}
\ee
Accordingly in regions occupied by the fluid  $(n\neq 0)$, $u^\alpha$ is normalized like any other 4-velocity.

Variation of $S$ with respect to $\kappa$ as well as $b^\b$ gives
\bea
u^\alpha \,b_\alpha = 0; \label{aux}
\\
F_{\alpha\beta}\,u^\a +\kappa\, u_\b=0. \label{preMHD}
\eea
Contracting Eq.~(\ref{preMHD}) with  $u^\b$ gives $\kappa\,
u^\b\,u_\b=0$; comparing with Eq.~(\ref{norm}) we conclude that
wherever $n\neq 0$, $\kappa=0$.  Thus Eq.~(\ref{preMHD}) is
equivalent to the MHD condition \reff{MHD}.

Variation of $S$ with respect to $u^\a$ gives
\be
{1\over 4\pi} F_{\alpha\beta}\, b^\b- \xi\, n\, u_\a- \zeta\,
n\,\eta_\a=0. \label{comp}
\ee
Contracting this with $u^\a$ and using  Eqs.~(\ref{MHD}) and
(\ref{norm})  leaves us with
\be
n\,(\xi-\zeta\,\eta_\a\, u^\a)=0. \label{sub}
\ee
On the other hand, we may vary $S$ with respect to $n$ to obtain
\be
n\,[\sfr12\,\xi\,(u^\a\, u_\a+1)+ \zeta\, \eta_\a\, u^\a]=0.
\ee
Comparison with Eq.~(\ref{sub}) shows that wherever $n\neq 0$ we
must have $ \xi=0$ as a consequence of Eq.~\reff{norm}. Combining
with our previous result on $\kappa$, we have for the local
Lagrange multipliers
 \be
 \kappa=\xi=0 \quad {\rm where}\quad n\neq 0.
 \ee
Now solving Eq.~(\ref{comp}) for $\eta_\a$ yields
 \be
 \eta_\a = {F_{\alpha\beta}\, b^\b\over 4\pi\zeta\, n},
 \label{eta}
 \ee
 from which it is obvious that (we  shall prove presently that $\zeta\neq 0$)
 \be
 \eta_\beta\, u^\beta= \eta_\beta\, b^\beta=0 \quad {\rm where}\quad n\neq 0.
 \label{etau}
 \ee

One further constraint is obeyed by $\eta_\beta$.   By the MHD condition \reff{MHD}  $F_{\alpha\beta}$ as given by Eq.~\reff{F} can be cast as
 \be
 F_{\alpha\beta}=\varepsilon_{\alpha\beta\gamma\delta}\, u^\gamma\,
 B^\delta.
 \label{FB}
 \ee
Using this form in Eq.~(\ref{eta}) shows that
\be
\eta_\beta\, B^\beta = 0 \quad {\rm where}\quad n\neq 0.
\label{etaB}
\ee
We see that the ``residual gauge field'' $\eta_\alpha$ is heavily constrained [Eqs.~(\ref{etau}) and (\ref{etaB})].

\subsection{Equations for scalar field}\label{scalareqs}

We shall now vary $S$ with respect to $\eta_\alpha$ obtaining
\be
n\,u^\alpha=\zeta^{-1}\,\sigma^2\,(\varphi,^\alpha-\eta^\alpha);
\label{velocity}
\ee
this equation links the hydrodynamical variables $n,
u^\alpha$ with the scalar field variables $\varphi, \sigma$.  If we
now vary with respect to $\varphi$ we get
\be \label{varphi}
[\sigma^2\, (\varphi,^\alpha-\eta^\alpha)\,(-g)^{1/2}],_\alpha =0.
\ee
Hence we see that the baryon current  density $nu^\alpha$ is
conserved,
\be
 (nu^\alpha)_{;\alpha}=0,
 \label{ncons}
\ee
as it should be.

Varying $S$ with respect to $\sigma$ gives in turn (a prime
signifies derivative with respect to $\sigma^2$)
\be
\sigma,_\alpha{}^{;\alpha} - \sigma\,(\varphi,^\alpha-\eta^\alpha)
(\varphi,_\alpha-\eta_\alpha)-\sigma\,V'(\sigma^2)=0. \label{sigma}
\ee
This last equations provides  a bridge to the thermodynamic
variables.  As in the case of the superfluid, which emerges from
complex scalar field dynamics only for long wavelengths, so here.
We imagine that whereas the phase $\varphi$ may be rapidly varying,
the amplitude $\sigma$ is slowly varying. More precisely, if $L$
denotes the scale of $\sigma$ variation, we assume
\be
L^{-2}\ll |(\varphi,^\alpha-\eta^\alpha)(\varphi,_\alpha-\eta_\alpha)|.
\label{L^2}
\ee
But  $\sigma,_\alpha{}^{;\alpha}/\sigma $  is then ${\cal
O}(L^{-2})$, so it must be negligible  compared to
$(\varphi,^\alpha-\eta^\alpha)(\varphi,_\alpha-\eta_\alpha)$. We
thus have
\be
(\varphi,^\alpha-\eta^\alpha)(\varphi,_\alpha-\eta_\alpha)=-\,V'(\sigma^2).
\label{square}
\ee
With the help of Eqs.~(\ref{velocity}) and (\ref{norm}) this can be written
\be
\zeta^2\, n^2 = \sigma^4 V'(\sigma^2). \label{sup}
\ee
from which it is clear that we must require $V'\geq 0$;  regions of
spacetime where this is not true cannot be occupied by the fluid.

\subsection{Electric current and Lorentz force}

If we vary $S$ with respect to $A_\alpha$ we obtain
\be
\left[\Big({1\over 4\pi}F^{\alpha\beta}- {1\over 2\pi}\,u^{[\alpha}
b^{\beta]}\Big)(-g)^{1/2}\right]_{,\beta}=0,
\ee
or more conveniently
\be
\Big(F^{\alpha\beta}-2\, u^{[\alpha} b^{\beta]}\Big)_{;\beta}=0.
\label{divF}
\ee
Comparing this with Maxwell's equations
$F^{\alpha\beta}{}_{;\beta}=4\pi\, J^\alpha$, where $J^\alpha$ is
the electric 4-current density, gives us
\be
J^\alpha=\frac{1}{2\pi}\big( u^{[\alpha} b^{\beta]}\big)_{;\beta}\  .
\ee
Carrying out the derivatives on suitably factored  quantities and
taking cognizance of Eq.~(\ref{ncons}) we have
\be
4\pi\,J^\alpha = \bra{n\,u^\alpha}_{;\beta}\bra{{b^\beta\over n}} +
\bra{{b^\beta\over n}}_{;\beta}\,n\,u^\alpha -\bra{{b^\alpha\over
n}}_{;\beta}\,n\,u^\beta.
\ee
Finally in view of Eq.~(\ref{MHD})  we have for the Lorentz force
density
\be
4\pi\,F_{\alpha\beta}\,
J^\beta=-F_{\alpha\beta}\,\left[\Big({b^\beta\over
n}\Big)_{;\gamma}\,nu^\gamma-(nu^\beta)_{;\gamma}\Big({b^\gamma\over
n}\Big)\right] .\label{Lorentz}
\ee
It is perhaps significant that the   quantity in the square brackets
is the Lie derivative of $b^\alpha/n$ along $nu^\beta$.

With an eye on reconstructing the Euler equation from our approach, let us use
this result to write the Lorentz force in an alternative form.  We first define
\be
{\cal F}_{\alpha\beta}\equiv \eta_{\beta;\alpha} - \eta_{\alpha;\beta}.
\label{calF}
\ee
We now substitute Eq.~(\ref{eta}) for $\eta^\alpha$ here
and carry out the derivatives:
\be
4\pi\,\zeta\,{\cal F}_{\beta\alpha}=
(F_{\a\g;\beta}-F_{\b\g;\alpha}){b^\gamma\over
n}+F_{\alpha\g}\,\Big({b^\gamma\over n}\Big)_{;\beta}
-F_{\beta\g}\,\Big({b^\gamma\over n}\Big)_{;\alpha}.
\ee
By the Maxwell equations
\be
F_{\alpha\beta;\gamma}+F_{\gamma\alpha;\beta}+F_{\beta\gamma;\alpha}=0
\label{maxwellh}
\ee
we may replace $F_{\alpha\g;\beta}-F_{\beta\g;\alpha}$
in the previous  equation
by $F_{\alpha\beta;\gamma}$.
This done, let us interchange the $\alpha$ and $\beta$ indices, and
contract the resulting expression with ${nu^\beta}$.  After
completing a derivative we have
\bea
4\pi\,\zeta\,{\cal F}_{\alpha\beta}\,nu^\beta &=&
-\big(F_{\alpha\beta}\, n u^\beta \big)_{;\gamma} {b^\gamma\over n}
+ F_{\alpha\beta}\,( n u^\beta )_{;\gamma}\,{b^\gamma\over n}
\nonumber
\\
&+&nu^\beta\bras{F_{\gamma\alpha}\Big({b^\gamma\over
n}\Big)_{;\beta} -F_{\gamma\beta}\,\Big({b^\gamma\over
n}\Big)_{;\alpha}}\!.
\eea
By virtue of the MHD condition (\ref{MHD}) the first  and fourth
terms on the r.h.s. of this last equation drop out.  By comparing
with Eq.~(\ref{Lorentz}) we see that
\be \label{f}
F_{\alpha\beta}\, J^\beta = \zeta \,{\cal F}_{\alpha\beta}\,
nu^\beta.
\label{FcalF}
\ee
Armed with this identity we turn to the recovery of Euler's MHD equation.

\subsection{The MHD Euler equation}\label{sec:Euler}

Let us introduce the notation
\be
\nu\equiv ({V'})^{1/2},
\label{musqrt}
\ee
where we explicitly mean the positive root.
In view of Eqs.~(\ref{sup}) and (\ref{velocity}) we now have
\bea
 n/\nu &=& \zeta^{-1}\, \sigma^2,
 \label{novermu}
 \\
\nu\, u^\alpha &=& \varphi,^\alpha-\eta^\alpha. \label{velocity2}
\eea
We now take the gradient of Eq.~(\ref{square})  and make use of
(\ref{velocity2}):
\be
\nu\, u^\alpha\,(\varphi_{;\alpha\beta}-\eta_{\alpha;\beta})=
-\nu\,\nu,_{\beta}.
\label{neweq}
\ee

We would here like to interchange the $\alpha$ and $\beta$ derivatives of
$\varphi$; however, we take note that the tensor
\be
W_{\a\b}\equiv \varphi_{;\b\a}-\varphi_{;\a\b}
\label{W}
\ee
need not vanish identically, as we might suppose,  because
$\varphi$ is a modular variable, and hence can be singular on various curves.
We may thus rewrite Eq.~\reff{neweq} as [see Eq.~\reff{calF}]
\be
\nu\, u^\alpha\,(\varphi_{;\beta\alpha}-W_{\a\b}-\eta_{\beta;\alpha}+ {\cal
F}_{\alpha\beta})=-\nu\,\nu,_{\beta}. \label{motion}
\ee
Dividing through by $\nu$, using (\ref{velocity2})  again, and
observing that $u^\alpha u_{\beta;\alpha} = Du_\beta/d\tau$ where
$\tau$ is the proper time of the observer with 4-velocity
$u^\alpha$, we get
\be
{Du_\beta\over d\tau}=-{\nu,_\beta\over  \nu}-u_\beta\, u^\alpha\,
{\nu,_\alpha\over \nu}+{({\cal F}_{\beta\alpha}-W_{\a\b}) \,u^\alpha\over  \nu
}, \label{preEuler}
\ee
which has the form of a fluid's equation of motion.  This has to be compared with the MHD Euler equation which has the form
\be
{Du_\beta\over d\tau}=-{p,_\beta\over \rho+p}- u_\beta\, u^\alpha
{p,_\alpha\over \rho+p} +{F_{\beta\alpha}\, J^\alpha\over \rho+p}.
\label{Euler}
\ee

First we shall work in a region outside singular curves of $\varphi$ where $W_{\a\b}=0$.  
The non-magnetic force terms on the r.h.s. of Euler's equation all
contain the pressure gradient in a particular way.
We observe that Eq.~\reff{preEuler} has terms with the same
structure involving the gradient of $\nu$.
This suggest that the two sets are identical.  But this can be true
only if
\be
{dp\over \rho+p} = {d\nu\over \nu}, \label{dp}
\ee
with the thermodynamic differentials taken at constant entropy.   On the other
hand, from thermodynamics we know that
\be
{d\rho\over dn} = {\rho+p\over n}, \label{drhodn}
\ee
where $n$, the baryon proper number density,  is the same quantity
we have been using above.  Using the last two equations we have
\be
{d(\rho+p)\over dn} = {\rho+p\over n} +{ dp\over d\nu}\,{d\nu\over
dn}= {\rho+p\over n}+  {\rho+p\over \nu}\,{d\nu\over dn}.
\ee
Dividing through by $(\rho+p)/dn$ reduces this to
\be
d(\rho+p)/(\rho+p) = dn/n+d\nu/\nu,
\ee
with integral
\be
\nu=P(\rho+p)/n=P\mu.
\label{nu=Cmu}
\ee
($P$ is an obviously positive constant of integration).
Thus  if our guess is correct, $\nu$  is proportional to
the relativistic enthalpy per baryon of Eq.~\reff{mu}.

Of course the ${\cal F}_{\a\b}$ term in
Eq.~\reff{preEuler} should be equivalent to the $F_{\a\b}$ term in Eq.~\reff{Euler}.
In light of result \reff{nu=Cmu} this is consistent with the identity \reff{FcalF} provided
 \be
P=\zeta^{-1}.
\label{ze}
\ee
We shall thus henceforth write $\mu=\zeta\nu$.
In view of the above we may conclude that  in
the region outside singular curves, Eq.~\reff{preEuler} from the field
formalism coincides with the MHD Euler equation.

Does this conclusion hold on the singular curves ?  That would
entail establishing that the constraint
\be
W_{\a\b}\,u^\b=0
\label{Wu=0}
\ee
holds on such curves.  We shall indeed prove this last condition in Sec.~\ref{kinematical} below. But for now let us note that formally Eq.~\reff{preEuler} is an integrability condition for Eq.~\reff{velocity2}.  First we notice that from the definitions of acceleration and ${\cal F}_{\b\a}$,
\bea
{Du_\beta\over d\tau}+{\nu,_\beta\over  \nu}+u_\beta\, u^\alpha\,
{\nu,_\alpha\over \nu}-{{\cal F}_{\beta\alpha}\,u^\alpha\over  \nu}
\\
=2{\big[(\nu u_{[\alpha}{})_{;\beta]} +\eta_{[\alpha}{}_{;\beta]}\big]u^\beta\over \nu}.
\eea
But according to Eq.~\reff{velocity2} this is equal to $-W_{\a\b}\, u^\a/\nu$.  Hence Eq.~\reff{preEuler} is satisfied identically.  Actually this is not surprising; the method used here in deriving it involves ``taking the curl'' of the expression for $u_\a, $ Eq.~\reff{velocity2}. 

\subsection{Gauge symmetry in the theory}\label{symmetry}

The vanishing of the divergence (\ref{divF})  means the quantity in
brackets is the dual of a 4-curl.  More precisely,
\be
F^{\alpha\beta}-2 \, u^{[\alpha} b^{\beta]}=\,
^*\!{f}^{\alpha\beta}, \label{Fandf}
\ee
where the antisymmetric tensor $f_{\a\b}$ and its dual
$^*\!{f}^{\a\b}$ are defined in terms of a field $a_\a$:
\bea
f_{\alpha\beta}&\equiv &a_{\beta,\alpha}-a_{\alpha,\beta},
\\
^*\!{f}^{\alpha\beta}& \equiv &{\scriptstyle 1\over\scriptstyle
2}\
\varepsilon^{\alpha\beta\gamma\delta}\,f_{\gamma\delta}=\varepsilon^{\alpha\beta\gamma\delta}\,
a_{\delta,\gamma}.
\eea
Let us take the dual of Eq.~(\ref{Fandf}):
\be
^*\!{F}_{\alpha\beta} -
\,\varepsilon_{\alpha\beta\gamma\delta}\,u^\gamma\,b^\delta = -
f_{\alpha\beta} \label{dualF}
\ee
As with any antisymmetric tensor,  we can relate $f_{\alpha\beta}$
to electric and magnetic-like vectors defined with respect to $u^\beta$, namely $e_\alpha$ and $b_\alpha$ given by
 \bea
 e_\alpha=f_{\alpha\beta}\,u^\beta,
 \label{e}
 \\
  b_\alpha= -^*\!{f}_{\alpha\beta}\,u^\beta.
\label{b}
\eea
Contracting Eqs.~(\ref{Fandf}) with $u_\beta$  shows that $b^\alpha$,
as just defined, is identical with the Lagrange multiplier
$b^\alpha$.  And contracting Eq.~(\ref{dualF}) with $u^\beta$ shows
that
\be
e_\alpha=B_\alpha.
\ee
We thus find that whereas $F_{\alpha\beta}$  comprises only a
magnetic part, $f_{\alpha\beta}$ comprises both electric and
magnetic parts, the first copying the magnetic part of
$F_{\alpha\beta}$.

The theory as stated has a $U(1)\times U(1)$  symmetry.  The action
and equations are invariant under
\bea
A_\alpha\rightarrow A_\alpha +\lambda^{(1)}{}_{,\alpha}\,,
\\
a_\alpha\rightarrow a_\alpha +\lambda^{(2)}{}_{,\alpha}\,,
\eea
with $\lambda^{(1)}$ and $\lambda^{(2)}$ two  independent functions.
This symmetry survives in the equations of motion after application
of the constraints.

We observe that the action $S_\psi$ is invariant under the transformation
\bea
\varphi&\rightarrow& \varphi +\Lambda\,,
\\
\eta_\alpha&\rightarrow& \eta_\alpha +\Lambda_{,\alpha}\,,
\label{changeeta}
\eea
for arbitrary $\Lambda$. However, this gauge invariance is
explicitly broken  by the $S_\zeta$ part of the action.  It is true
that once one takes the conservation law (\ref{ncons}) into account,
any change of $S_\zeta$ induced by transformation (\ref{changeeta})
can be converted into a surface term by means of Gauss' theorem.
But the use of equations of motion derived from the action is
inappropriate at a stage where symmetries of the action are being
considered.  Thus the $U(1)\times U(1)$ symmetry is all there is.

\section{Character of the MHD fluid}
\label{character}

There are two ways to bring out the character of the  MHD fluid in
the theory.  One is to exploit Eq.~(\ref{mu}) and its logical predecessors, all of which follow from differentiation of Eq.~(\ref{square}).
This will be done in Sec.~\ref{state}.  The second way is to
formally construct the energy-momentum tensor as it would occur in
Einstein's equations.  This road will be travelled in
Sec.~\ref{Tmunu}.

\subsection{Equation of state}\label{state}

As with any scalar field representation of a fluid,  we may
establish the relation between the form of $V(\sigma^2)$ and the
equation of state.  In light of Eqs.~\reff{musqrt} and \reff{ze}, we first write
Eq.~(\ref{drhodn}) in the form
\be
{d\rho\over d\sigma^2} = \mu\, {dn \over d\sigma^2} =\zeta\,
\sqrt{V'(\sigma^2)}\, {dn \over d\sigma^2}\,. \label{drhodsigma}
\ee
Then from Eq.~(\ref{novermu}) we obtain
\be
\zeta{dn\over d\sigma^2} = [V'(\sigma^2)
]^{1/2} +{\scriptstyle 1\over \scriptstyle  2} \sigma^2 V''(\sigma^2)
 [V'(\sigma^2)
]^{-1/2}
\ee
so that (\ref{drhodsigma}) becomes
\be
{d\rho\over d\sigma^2} = [V'(\sigma^2)
+{\scriptstyle 1\over \scriptstyle  2} \sigma^2 V''(\sigma^2)]\,.
  \label{drhodsigma2}
\ee
Integrating this, and then integrating by parts leads to
\be
\rho={\scriptstyle 1\over \scriptstyle  2}\,[V(\sigma^2)
+\sigma^2\, V'(\sigma^2)]\,.  \label{rho}
\ee

We now need an analogous expression for $p$.   Substituting
Eq.~(\ref{mu})  in Eq.~(\ref{dp}) allows us to write
\be
{dp\over d\sigma^2}= \zeta n\,{d\nu\over d\sigma^2}\,.
\ee
Recalling that
$\nu=(V')^{1/2}$, carrying out the differentiation, and substituting here from Eq.~(\ref{novermu}) we have
\be
{dp\over d\sigma^2}={\scriptstyle 1\over \scriptstyle  2}\,
\sigma^2 V''(\sigma^2)\,.
\label{dpdsigma2}
\ee
Integration followed by integration by parts leads to
\be
p={\scriptstyle 1\over \scriptstyle  2}\,[\sigma^2\, V'(\sigma^2)-V(\sigma^2)]\,.
\label{p}
\ee

Together with Eq.~(\ref{rho}) this equation  provides the equation
of state of the fluid in parametric form.  What are the general
requirements on $V(\sigma^2)$ which yield a physically acceptable
equation of state ?    One emerges from the need to
enforce causality, specifically that the squared speed of purely
acoustic perturbations, $v_s{}^2$, be positive and less than unity.
Dividing Eq.~(\ref{dpdsigma2}) by (\ref{drhodsigma2}) gives
\be
v_s{}^2 ={dp\over d\rho}= {V''\over V''+2V'/\sigma^2}.
\ee
Since we have already required $V'>0$, we must have that $V''>0$ as well. These two conditions automatically insure that $\rho>0$.  Requiring $p>0$  is
perhaps too strong a condition: there is also electromagnetic
pressure which may counter instability from negative fluid pressure.
But it is safe to require that the strong energy condition be
satisfied: $\rho+3 p>0$.  From expressions (\ref{rho}) and (\ref{p})
this gives $V<2\sigma^2 V'$, i.e., the logarithmic slope of
$V(\sigma^2)$ must be larger than $1/2$.  Thus the potential
\be
V(\sigma^2)=\Sigma_{k=1}^K\ A_{2k}\, \sigma^{2k},
\ee
with nonnegative coefficients $A_{2k}$ and finite $K$, satisfies these conditions.

\paragraph*{Example 1:} Assume $V=(m/\zeta)^2 \sigma^2$ ($m$ constant).
This corresponds to 
$p=0$, $\mu=m$  and $\rho=m\,n$, and 
represents a pressureless gas whose baryons have rest mass
$m$.  

\paragraph*{Example 2:} Take
$V=\beta\sigma^4$.  This gives $p={\scriptstyle 1\over\scriptstyle
3}\,\rho$ while  $\rho=3\cdot 2^{-5/3}\zeta^{4/3} \beta^{1/3} n^{4/3}$ and
$\mu=\zeta\sqrt{2\beta}\,\sigma$.  This represents  thermal radiation
with a sprinkling of baryons: the proportionality $\rho\propto
n^{4/3}$ is typical of adiabatic compression of radiation.

\subsection{Energy-momentum tensor}\label{Tmunu}

Einstein's equations are derived by varying the  total action
(including the Einstein-Hilbert part) with respect to
$g^{\alpha\beta}$.  From this follows that the energy-momentum
tensor $ T_{\alpha\beta}$ that sources these equations is determined
by
\be
\delta S = -  {\scriptstyle 1\over\scriptstyle 2}\int
T_{\alpha\beta}\,(-g)^{1/2}\,\delta g^{\alpha\beta}\, d^4x.
\ee
where the variation is to be  carried out before any of the
constraints or equations of motion found earlier are enforced.

Let us first carry out the  variation of $S_c$.  Because $A_\alpha,
b^\alpha$ and $u^\alpha$ are regarded as fundamental variables, the only contribution of the first term comes from the $g^{\alpha\beta}$
dependence of $(-g)^{1/2}$.  This will give rise to a
$g_{\alpha\beta}$ factor which will multiply $F_{\gamma\delta}\,
b^\delta u^\gamma$.  However, once we take Eq.~(\ref{MHD}) into
account this term vanishes.  The second term in $S_c$ depends on the
metric in two ways; however, it is all multiplied by $\kappa$
which we know will vanish.  Hence $S_c$ contributes nothing to the
energy-momentum tensor.  Strictly speaking the above is true only in
the region occupied by the fluid, that where $n\neq 0$.  For without
this condition we are unable to derive the MHD condition, or the
condition $\kappa=0$.  So we should say that $S_c$ contributes
nothing to the energy-momentum tensor of the fluid.  A similar
remark will apply below.

Variation of $S_{nor}$ with respect  to $g^{\alpha\beta}$ produces
two types of terms, both of which are multiplied by $\xi$, which we
know will vanish.  Hence $S_{nor}$ also does not contribute to the
energy-momentum tensor of the fluid.

Turning to $S_\zeta$ we meet a similar  situation.  $u^\alpha$ and
$\eta_\beta$ are fundamental variables, so no metric enters in their
scalar product.  Variation of $g_{\a\b}$ in $(-g)^{1/2}$  contributes a
factor $g^{\alpha\beta}(-g)^{1/2} $ which is multiplied by the $u^\gamma\eta_\gamma$.  Since this last term ultimately vanishes by Eq.~(\ref{etau}), $S_\zeta$ too makes no contribution to the fluid's energy-momentum tensor.

It is plain that variation of $S_M$ with  respect to the metric will
produce the Maxwell energy-momentum tensor.  Therefore, the MHD
fluid energy-momentum tensor comes exclusively from $S_\psi$.
Carrying out the variation gives us the sum
$T^{(\sigma)}_{\alpha\beta}+T^{(\varphi)}_{\alpha\beta}$, where
\bea
T^{(\sigma)}_{\alpha\beta}&\equiv& \sigma,_\alpha \sigma,_\beta  -
{\scriptstyle 1\over\scriptstyle 2} \sigma,_\gamma \sigma,^\gamma\,
g_{\alpha\beta},
\label{sigmaT}
\\
T^{(\varphi)}_{\alpha\beta}&\equiv&
\sigma^2\,(\varphi,_\alpha-\eta_\alpha) (\varphi,_\beta-\eta_\beta)
\label{phiT}
\\
\nonumber
& -& {\scriptstyle 1\over\scriptstyle
2}\big[\sigma^2(\varphi,_\gamma-\eta_\gamma)
(\varphi,^\gamma-\eta^\gamma) +V(\sigma^2)\big]\, g_{\alpha\beta}.
\eea

Let us look at $T^{(\sigma)}_{\alpha\beta}$.  The  first term is
${\cal O}(\sigma^2\, L^{-2})$  in the terminology of
Sec.~\ref{scalar}.  According to inequality (\ref{L^2}),  that term
is negligible compared to the first term in
$T^{(\varphi)}_{\alpha\beta}$.  By the same logic $\sigma,_\gamma
\sigma,^\gamma$ may be neglected compared to
$\sigma^2(\varphi,_\gamma-\eta_\gamma)
(\varphi,^\gamma-\eta^\gamma)$.  We thus see that  the fluid's
energy-momentum tensor is dominated by
$T^{(\varphi)}_{\alpha\beta}$.  Let us replace in this last
$\varphi,^\a-\eta^\a$ by its expression (\ref{velocity2}) and simplify by means
of Eqs.~(\ref{musqrt})-(\ref{novermu}).  The result is
 \be
T^{(\varphi)}_{\alpha\beta} =\sigma^2\,V'(\sigma^2)\,u_\alpha\,
u_\beta+{\scriptstyle 1\over\scriptstyle 2} \big[
\sigma^2\,V'(\sigma^2)-V(\sigma^2) \big]g_{\alpha\beta}.
 \ee
 This may be compared with the standard perfect fluid energy-momentum tensor
 \be
 T^{(f)}_{\alpha\beta}=(\rho+p)u_\alpha\, u_\beta+p\,g_{\alpha\beta}.
 \ee
The $\rho$ and $p$ may now be identified; they coincide with those
given by Eqs.~(\ref{rho}) and (\ref{p}).  The pictures obtained from
the field equations and from the energy momentum tensor are thus
consistent.

\section{The 1+3 viewpoint}
\label{3+1}

\subsection{Preliminaries}

As we have seen, our proposed field theory for MHD leads naturally
to the fluid 4-velocity
\be
u^\a=\bra{\varphi^{,\a}-\eta^\a}/\nu\,,  \label{4vel}
\ee
and it gives rise to the projection tensor $h_{\a\b} \equiv
g_{\a\b}+u_\a u_\b$ projecting orthogonally to $u^\a$. Decomposing
the covariant equations in parts parallel and normal to $u^\a$ will
render the effects of curvature more transparent and enhance our
physical understanding. We adopt in the following the notation of Ref.~\onlinecite{Ellis-vanElst} for convenience.

There are now two differentiation operators,  namely the (proper)
time derivative ${\cal D} \equiv u^\a\nabla_\a $ (often denoted by
an overdot: ${\cal D}f\equiv \dot f$) and the (totally projected)
spatial derivative $\D^\a$ (e.g., $\D^\a\,T^{~\g}_{\b}=
h^{\a\s}\,h_{\b}^{~\d}\,h^{\g}_{~\rho}\,\nabla_\s\,T^{~\rho}_{\d})$,
respectively.
The covariant derivative of \emph{any} scalar function $f$ is thus decomposed as
$\nabla^\a f = \D^\a f - u^\a{\cal D} f $. A congruence of observers with
4-velocity $u^\a$ is covariantly characterized in terms of the
associated Raychaudhuri~\cite{R} kinematical variables, which are derived from the
covariant derivative of $u^\a$:
\bea 
 \nabla_{\a}u_{\b} &=& -\,u_\a\,\dot{u}_\b + \D_{\a}u_{\b}\,,
 \label{kq} 
 \\
\D_{\a}u_{\b} &=& {\sfr13}\,\Theta\,h_{\a\b} + \sigma_{\a\b} +
\omega_{\a\b} \,.
\label{Du}
\eea
Here the trace $\Theta \equiv \D_\a u^\a{}$ is the {\em volume
rate of expansion\/} of the congruence; $\sigma_{\a\b} \equiv
\D_{(\a} u_{\beta)} -\sfr13\Theta h_{\a\b}$ is
the trace-free symmetric {\em rate of shear\/} tensor
($\sigma_{\a\b} = \sigma_{(\a\b)}$, $\sigma_{\a\b}\,u^\b = 0$,
$\sigma^\a{}_{\a}= 0$), describing the rate of distortion of the
congruence; and $\omega_{\a\b} \equiv \D_{[\a}u_{\b]}$ is the
skew-symmetric {\em vorticity\/} tensor ($\omega_{\a\b} =
\omega_{[\a\b]}$, $\omega_{\a\b}\,u^\b = 0$), describing the
rotation of the congruence relative to a non-rotating
(Fermi-Walker propagated) frame.  Finally ${\cal D}u^\a = \dot u^\a= u^\b\nabla_\b u^\a$
is the {\em relativistic acceleration\/} vector, which represents
the influence of forces other than gravity on the observer (a
free-falling observer has vanishing acceleration in her rest-frame).

A crucial feature of the newly introduced derivative operators is
that they do not commute with each other in general. In particular,
one finds for any scalar function $f$ the commutation relations
\bea
\D_{[\a}\D_{\b]}f &=& \omega_{\a\b}\,\dot f, \label{com1}
\\
\D_\a\dot f -h_\a^{~~\b}\,{\cal D} \bra{\D_\b f} &=& -\dot u_a\, \dot f
+ (\D_\a u^\b)\, \D_\b f\ .\label{com2}
\eea
Eq.~\reff{com1} immediately tells us that the spatial derivative
$\D$ is a covariant derivative if and only if the vorticity
$\omega_{\a\b}$ vanishes, in which case the 4-velocity $u^\a$ is
hypersurface-orthogonal and  $h_{\a\b}$ becomes the induced metric
of the hypersurface.

It is convenient to introduce the totally antisymmetric tensor
(spatial volume element) $\e_{\a\b\g} \equiv u^\d\varepsilon_{\d\a\b\g}$.
With its help we define the spatial $\curl$ of a 3-vector $V^\a$
$(V^\a u_\a=0)$ to be $\curl V^\a \equiv \e^{\a\b\g}\D_\b V_\g$.
Moreover, we define the vorticity vector $\omega^\a \equiv \sfr12
\e^{\a\b\g}\omega_{\b\g}$ such that
$\omega_{\a\b}=\e_{\a\b\g}\omega^\g$.  Analogously,
$B^\a \equiv \sfr12
\e^{\a\b\g}F_{\b\g}$  and $F_{\a\b}=\e_{\a\b\g}B^\g$.

We note that the above commutation relations have to be modified for the case of the modular field $\varphi$, since its curl $W_{\a\b}$ is not vanishing. Decomposing $W_{\a\b}$ into electric and magnetic components,
\be
W_{\a\b}=u_\a\,W_\b-u_\b\,W_\a + \epsilon_{\a\b\g}V^\g\ ,
\label{wdec}
\ee
the modified commutation relations \reff{com1} and \reff{com2} can then be written as
\be
\D_{[\a}\D_{\b]}\varphi = \epsilon_{\a\b\g}\bra{\omega^{\g}\,\dot \varphi + \sfr12\,V^\g} \label{com1m}
\ee
and
\be
\D_\a\dot \varphi -h_\a^{~~\b}\,{\cal D} \bra{\D_\b \varphi} = -\dot u_a\, \dot \varphi
+ (\D_\a u^\b)\, \D_\b \varphi + W_\a\ ,\label{com2m}
\ee
respectively. Note that in this section we do not enforce the constraint~\reff{Wu=0}, $W^\a=W^{\a\b}u_\b=0$, in order to demonstrate which findings hold irrespective of it.

\subsection{Kinematical variables}
\label{kinematical}

We now determine the kinematical variables associated with a
congruence of observers with 4-velocity $u^\a$ given by \reff{4vel}.
First by contracting \reff{4vel} with $u_\a$ and
$h_{\a\b}$, respectively, we get
\bea
\nu &=& - \dot\varphi\,,
\label{mu1}
\\
\eta^\a &=& \D^\a \varphi\,,\label{eta1}
\eea
Likewise, from Eq.~\reff{novermu} we obtain
\be
n = -\zeta^{-1}\,\s^2\dot\varphi\,.
\label{n1}
\ee
Hence the number density $n$ is determined in terms of both
$\s$ and $\dot\varphi$, while the field $\eta^\a$ is
determined by the spatial dependence of $\varphi$ alone.

The expansion of the congruence is most easily obtained by writing out the divergence in Eq.~\reff{ncons} and realizing that $\D_\a u^\a=\nabla_\a u^a$:
\be
\Theta = -\frac{\dot n}{n}\,.\label{theta}
\ee

The congruence $u^\alpha$ has non-vanishing vorticity. This is most readily
inferred by making use of the modified commutator relation \reff{com1m}
for $\varphi$  and also Eq.~\reff{eta1}, giving
\be
\omega_{\a\b} = \frac{1}{\dot\varphi}\bra{\D_{[\a}\,\eta_{\b]} -
\frac{1}{2}\,\epsilon_{\a\b\g}V^\g} .
\ee
Further, the
shear of the congruence is given by the expression
\be
\s_{\a\b} =
-\frac{1}{\dot\varphi}\,\D_{(\a}\big[\nabla_{\b)}\varphi-\eta_{\b)}\big] +
\frac13\, \frac{\dot n}{n}\,h_{\a\b}\,,
\ee
where we used Eq.~\reff{theta} to replace the expansion term.

It remains to calculate the 4-acceleration $\dot u^\a$. 
We use the Euler equation \reff{preEuler} for this task,
which we write in the form
\be
\nu\,\dot u^\a = -\D^\a \nu + \bra{\zeta\,n}^{-1}F^\a{}_\b J^\b-W^\a\,,
\label{Euler3}
\ee
where we have employed Eq.~\reff{f}. Writing $F_{\a\b}=\e_{\a\b\g}B^\g$ and replacing  $\nu$ via expression~\reff{mu1}, one gets
finally
\be
\dot u^\a = -\frac{1}{\dot\varphi}\,\bra{\D^\a\dot\varphi +
\frac{1}{\zeta\, n}\,\e^{\a\b\g}J_\b B_\g -W^\a}.
\ee
It is instructive to calculate $\dot u^\a$ in another way, namely by
employing the commutation relation \reff{com2m}  with
\reff{eta1}. Doing so one obtains
\be
\dot u^\a = -\frac{1}{\dot\varphi}\,\bra{\D^\a\dot\varphi - \bras{
\dot\eta^{\la\a\ra}+\eta_\b\D^\a u^\b}-W^\a} ,
\ee
where the angle brackets denote projection with $h_{\a\b}$.
Comparing the last two expressions for the 4-acceleration, we find
an evolution equation for the field $\eta^\a$, namely
\be
\dot\eta^{\la\a\ra}+ \sfr13\Theta\eta^\a
+\bra{\s^\a_{~~\b}+\omega^\a_{~~\b}}\eta^\b =
-\bra{\zeta\,n}^{-1}\e^{\a\b\g}J_\b B_\g . \label{Hall}
\ee
We thus see that the electromagnetic interaction induces spatial inhomogeneity in
the scalar field $\varphi$ via the Lorentz force $\mathbf J \times \mathbf B$.  The last equation will be useful in Sec.~\ref{sec:bernoulli}.

It may be worthwhile to investigate the relationship between the tensors ${\cal F}_{\a\b}=2\,\nabla_{[\a}\,\D_{\b]}\,\varphi$ and $W_{\a\b}=2\,\nabla_{[\a}\,\nabla_{\b]}\,\varphi$, respectively.   Expanding $\D_\b$ in the definition of ${\cal F}_{\a\b}$ in terms of $\nabla_\b$ we readily obtain
\be
{\cal F}_{\a\b} = W_{\a\b} - 2\,\nabla_{[\a}\bra{\nu\,u_{\b]}}.
\ee
This can be further rewritten by employing in the r.h.s. the decompositions \reff{kq}, \reff{wdec}, the commutator relations \reff{com1m}, \reff{com2m}, and the result \reff{Hall}. Some straightforward algebra reveals that
\be
{\cal F}_{\a\b} =  \frac{2}{\zeta\,n}\,u_{[\a}\,F_{\b]\g}\,J^\g + 2\,\D_{[\a}\,\eta_{\b]},
\label{la}
\ee
from which the electric and magnetic parts of ${\cal F}_{\a\b}$ are manifest.  The magnetic part may be written in an alternative way by means of Eq.~\reff{com1m}.   Eq.~\reff{la} is fully consistent with Eq.~\reff{FcalF}.

We are now in the position to give a derivation of the
constraint~\reff{Wu=0}, that is $W^\a =0$, by looking at overall energy-momentum
conservation. We recall from Sec.~\ref{Tmunu} that the
action for the complex scalar field $\psi$ yields a perfect fluid
energy-momentum tensor $T_{\a\b}^{(\varphi)}$ [cf. Eq.~\reff{phiT}]
in the long-wavelength limit, with the
number density $n$ and the potential $V$ related via Eq.~\reff{sup}. The
concomitant energy density and pressure of the fluid as given
in~\reff{rho} and \reff{p} can be recast as
\bea
\rho = {\scriptstyle 1\over \scriptstyle
2}\bras{\bra{\frac{\zeta\,n}{\sigma}}^2 + V(\sigma^2)} = {\scriptstyle
1\over \scriptstyle  2}\bras{\bra{\sigma\,\dot\varphi}^2 + V(\sigma^2)}
, \label{rho1}
\\
p = {\scriptstyle 1\over \scriptstyle
2}\bras{\bra{\frac{\zeta\,n}{\sigma}}^2 - V(\sigma^2)} = {\scriptstyle
1\over \scriptstyle  2}\bras{\bra{\sigma\,\dot\varphi}^2 - V(\sigma^2)},
\label{p1}\
\eea
where Eqs.~\reff{sup} and~\reff{n1} have been used. 

Now, the overall conservation of energy-momentum requires
$\nabla_{\b}T^{\a\b}_{(\varphi)} = F^{\a\b}\,J_{\b}$.  The
familiar energy and momentum conservation equations follow by
contracting this with $u_\a$ or by spatially projecting on the index
$\a$, respectively:
\bea
\dot\rho + \Theta\bra{\rho + p} &=& 0\,,\label{en1}
\\
\bra{\rho + p}\dot u^\a + \D^\a p &=&
\epsilon^{\a\b\g}\,J_\b\,B_\g\,.\label{mom1}
\eea
We can check the consistency of these equations by inserting the
expressions \reff{rho1} and \reff{p1}. Observing that $\dot V =
2\,\sigma\,\dot\sigma V'$ (with an analogous relation for $\D^\a V$) and
remembering Eq.~\reff{theta}, the energy equation~\reff{en1} is readily
seen to hold. Now, by calculating the spatial gradient of
the pressure, $\D^\a p$, taking the commutator relation~\reff{com2} as
well as Eq.~\reff{Hall} into account, and comparing the result with the
momentum equation~\reff{mom1}, we end up with
\be
\zeta\,n\,W^\a = 0\ .
\ee
This means that, in regions occupied by the fluid, the
constraint~\reff{Wu=0} must hold in order to implement the modularity of
the scalar field $\varphi$  in a consistent manner.  Thus we have shown that the derivation of the MHD Euler equation from Eq.~\reff{square}, as carried out in Sec.~\ref{sec:Euler}, is also valid  on singular curves.  And, of course, the momentum conservation Eq.~\reff{mom1} is already the Euler equation, here obtained directly from energy-momentum conservation.

\subsection{The role of the Lagrange multiplier $b^\a$}

In order to get a better understanding of the ``little magnetic
field" $b_\a$, we look at Maxwell's equations from the 1+3
viewpoint. Due to our MHD condition of vanishing electric fields,
$E^\a =0$, these become now
\bea
4\pi\,J^\a &=& F^{\a\b}_{~~~;\b} = \bra{\e^{\a\b\g}\,B_\g}_{;\b} , \\
0 &=& ^*\!{F}^{\a\b}_{~~~;\b} = -2\bra{u^{[\a}\,B^{\b]}}_{;\b} .
\label{B}
\eea
As usual, the 4-current is decomposed relative to $u^\a$ in the
manner
\be
J^\a = \varrho\,u^\a + J^{\la \a \ra}; \quad J^{\la \a \ra}
= h^{a}_{~\b}\,J^\b,
\ee
where $\varrho=-J_\a u^\a$ is the charge density and
$J^{\la \a \ra}$ is the Maxwell 3-current, respectively. Projecting
the above Maxwell equations perpendicularly and along $u^\a$, one
arrives at the following system of equations:
\bea
4\pi\,J^{\la \a \ra} &=& \curl B^\a +\e^{\a\b\g}\,\dot u_\b\,B_\g,
\label{mj}
\\
4\pi\,\varrho &=& -2\,\omega_\a\,B^\a,\\
0 &=& \dot B^{\la \a \ra} + \sfr23\,\Theta\,B^\a -
\bra{\s^{\a}_{~~\b}-\omega^{\a}_{~~\b }}B^\b,
\label{dotB}
\\
0 &=& \D_\a\,B^\a.
\eea
These equations are just the familiar special relativistic Maxwell
equations with vanishing electric field $E^\a$ but with the general
relativistic correction terms encoded in the kinematical variables
of the observers' 4-velocity $u^\a$ (c.f. Ref.~\onlinecite{multifluids}, for example). On the other hand, in light of
Eqs.~\reff{Fandf} and \reff{dualF} Maxwell's equations may equally
well be written in the form
\bea
4\pi\,J^\a = F^{\a\b}_{~~~;\b} = 2\bra{u^{[\a}\,b^{\b]}}_{;\b}, \label{E} \\
0 = ^*\!{F}^{\a\b}_{~~~;\b} =
\bra{\e^{\a\b\g}\,b_\g-f^{\a\b}}_{;\b} =
-2\bra{u^{[\a}\,e^{\b]}}_{;\b} .\label{E1}
\eea
It is immediately clear from Eq.~\reff{E} that the Lagrange
multiplier $b^\a$ plays a role analogous to an electric field,
whilst Eq.~\reff{E1} is identically to Eq.~\reff{B} since we already
know that $e^\a = B^\a$.  Decomposing the inhomogeneous equation
\reff{E}  readily yields
\bea
4\pi\,J^{\la \a \ra} &=& -\dot b^{\la \a \ra} - \sfr23\,\Theta\,b^\a
+ \bra{\s^{\a}_{~~\b}-\omega^{\a}_{~~\b }}b^\b, \label{mj1}
\label{beq}
\\
4\pi\,\varrho &=& \D_\a\,b^\a,
\eea
which look like the Ampere-Gauss equations, and once again demonstrate the ``electric" nature of the Lagrange
multiplier $b^\a$.  Thus we might view $\zeta\, n\,\eta^\a =
(4\pi)^{-1}\,\e^{\a\b\g}\,b_\b\, B_\g$ as a Poynting vector.

It should be noted that the Lagrange multiplier $b^\a$ is anything
but independent. Suppose suitable scalar fields $\s$ and $\varphi$ were chosen, fixing $u^\a$ and the kinematical variables, and a solution for the magnetic field $B^\a$ had been found. Comparison of
Eqs.~\reff{mj} and \reff{mj1} then reveals that $b^\a$ can be
determined via a differential equation wherein $B^\a$ acts as a
source. Our model is thus consistent as long as that differential
equation possesses a non-trivial solution for $b^\a$.  Such a solution is not unique.  For according to Eq.~\reff{dotB}, a multiple of $B^\a$ may be added to $b^\a$ without changing the latter's status as a solution of Eq.~\reff{mj1}.  The addition to $b^\a$  of a multiple of $u^\a$, which is suggested as a possibility by Eq.~\reff{E}, is made untenable by Eq.~\reff{aux}.

\section{Circulations and helicities}
\label{hel}

\subsection{Conserved circulation}
\label{con-cir}

We return to Eq.~\reff{4vel}.  Solving for $\varphi_{,\alpha}$, substituting $\eta_\alpha$ from Eq.~\reff{eta}, and integrating the result over a closed loop initially lying on a spacelike surface and carried along by the flow, we obtain
\be
\Gamma_{O}\equiv \oint \left(\mu\,u_\alpha+\zeta\eta_\a \right) dx^\alpha=2\pi \zeta N,
\label{Gamma}
\ee
where $N\in \mathbb{Z}$ counts the number of times the phase $\varphi$ winds around its natural interval $[0,2\pi]$ as the point traverses  the loop once.  At a purely classical level this result gives conservation of the circulation $\Gamma_O$.  This is because hydrodynamic evolution amounts to a continuous deformation of the loop and a continuous variation of the integrand in $\Gamma_O$.  By continuity $N$ cannot change during such an evolution, and so $\Gamma_O$ is conserved.

The conservation of $\Gamma_O$ was put in evidence by Bekenstein and A. Oron~\cite{BAO}; they employed a purely hydrodynamic variational principle for MHD (in contrast to the present field-theoretic one) to generalize a circulation conservation law discovered by E. Oron and given in Ref.~\onlinecite{BEO} (henceforth BEO) for flow with both stationary and axial symmetry.  What is new in our result here is the quantization of the circulation, a direct result of the use of a phase as a dynamical variable.  By contrast, Ref.~\onlinecite{BAO} used only thermodynamic variables and the velocity to describe the fluid.

What is the meaning of the quantization ?  How does it square with the presumed continuity in allowed values of circulation ? We notice from Eq.~\reff{actionintegral} that the dimensions of $\zeta$ must be those of action.  This because $\eta_\alpha$ carries the same dimension as $\varphi_{,\alpha}$, namely those of reciprocal length, $L^{-1}$, while $n$ by definition has dimensions of  $L^{-3}$, and $u^\alpha$ is dimensionless ($c=1$).  Now let us consider a situation with arbitrarily weak magnetic field frozen into a nonrelativistic fluid.  According to Eq.~\reff{mu} the second condition means $\mu$ equals the rest mass $m$ per particle; and the spatial part of $u^\alpha$ reduces to the usual 3-velocity $\mathbf{v}$.  The condition \reff{Gamma} is thus
\be
\oint \mathbf{v}\cdot d\ell =N{2\pi\zeta\over m}.
\label{OF}
\ee
Now  our action \reff{S_phi}  in the limit of vanishing  magnetic terms, and with neglect of derivatives of $\sigma$, is suitable for describing a superfluid condensate.  However, for a superfluid circulation is quantized by the Onsager-Feynman rule~\cite{LLSP}, which is precisely \reff{OF} with $\zeta=\hbar$. We must thus calibrate $\zeta$ to a value $\hbar$ in all cases (although in deference to our classical approach we shall continue to use the notation $\zeta$).  The smallness of the quantum of action means that for macroscopic loops the Oron circulation $\Gamma_O$ will take on an almost continuous range of values, as would be expected classically.

The above approach is germane to a field theoretic approach; but how can one understand the conservation of $\Gamma_O$ at the macroscopic level ?  For this purpose we shall define the 4-vector
\be
w_\alpha=\nu u_\alpha +\eta_\alpha,
\label{w}
\ee
this being nothing else than $\zeta^{-1}$ times the vector constituting the integrand of Eq.~\reff{Gamma}.
The fact that $w_\alpha=\varphi_{,\alpha}$ does not make the vector's curl $W_{\a\b}$ [defined by Eq.~\reff{W}] identically zero because, as mentioned earlier,  $\varphi$ is a modular variable, and hence can be singular on various curves (these would be the cores of vortices of the spatial vector $h^{\a\b}\,w_\b=\eta^\a$).   As shown by Eq.~\reff{Wu=0},
\be
W_{\a\b}\,u^\b=0.
\label{W2}
\ee
An obvious identity satisfied by the curl $W_{\alpha\beta}$ is [see Sec.~\ref{intro}]
\be
W_{[\alpha\beta;\gamma]}=0\qquad {\rm or} \qquad ^*\! W^{\alpha\beta}{}_{;\beta}=0.
\label{p2}
\ee

We now consider the rate of change of Oron circulation, or more precisely, its change when the contour is Lie-dragged along $u^\alpha$.  As a first step we write by means of Stokes' theorem
\be
\Gamma_O=\zeta\oint W_{\alpha\beta}\, d\Sigma^{\alpha\beta},
\ee
with $ d\Sigma^{\alpha\beta}$ the 2-area element tensor on a surface spanning the loop in question.
Secondly, according to a calculation by Bekenstein and E. Oron (BEO)~\cite{BEO}, for \emph{any} tensor $W_{\alpha\beta}$ and any loop
\be
{\cal L}_u\,\Gamma_O=\zeta \oint \big[2 ( u^\gamma\, W_{\gamma[\beta})_{;\alpha]}-3W_{[\alpha\beta;\gamma]}\, u^\gamma \big]\, d\Sigma^{\alpha\beta}.
\label{Eli}
\ee
And of course, the integrand here vanishes by Eqs.~\reff{W2} and \reff{p2}.  Thus the circulation $\Gamma_O$ is conserved quite apart from the pertinent vector being the gradient of a phase.  Both because of this and because of its limiting form when $F_{\alpha\beta}\rightarrow 0$, we may regard $\Gamma_O$ as the generalization of Kelvin's circulation to MHD.

We recall that the magnetic (Alfven) circulation
\be
\Gamma_A=\oint A_\alpha dx^\alpha=
\int F_{\alpha\beta}\, d\Sigma^{\alpha\beta}
\ee
is also conserved in perfect MHD.  This may be shown exclusively from Eq.~\reff{Eli} with $W_{\alpha\beta}\to F_{\alpha\beta}$ and the conditions \reff{MHD} and \reff{maxwellh}.  In Sec.~\ref{C} we shall exhibit a third, new, circulation which is conserved in situations with two spacetime symmetries.

\subsection{Conserved helicities}

According to Moffatt~\cite{moffatt} the conservation of fluid helicity in perfect pure fluid dynamics reflects the conservation of Kelvin circulation around linked vortex lines.   Likewise, the conservation of the fluid-magnetic Woltjer helicity~\cite{woltjer} in MHD reflects the conservation of the Kelvin circulation around a vortex \emph{and} the Alfven circulations around a magnetic flux line linked with the vortex.  And  Woltjer magnetic helicity reflects the conservation of Alfven circulation around linked flux lines.  Having displayed the new circulation $\Gamma_O$, we may ask whether it furnishes new helicity conservation laws to replace the lost Moffatt helicity in the MHD regime.

For a perfect pure relativistic fluid the Moffatt helicity conserved current is~\cite{carter,bekhelicity}
\be
H_{\rm f}^\alpha = \sfr{1}{2}\varepsilon^{a\b\g\d}\,(\mu u_{[\d}{}
)_{;\g]}\,\mu u_\b=-\mu^2\omega^\a.
\ee
This may be compared with the conserved currents for Woltjer's magnetic and fluid-magnetic helicities in perfect MHD flow~\cite{carter,bekhelicity}:
\bea
H_{\rm m}^\a &=&^*\!{F}^{\a\b}\,A_\b=-B^\b{\cal A}_\b u^\a-\phi B^\a,
\label{Hm}
\\
H_{\rm fm}^\alpha &=&^*\!{F}^{\a\b}\,\,\mu u_\b=-\mu B^\a,
\eea
where $\phi\equiv -A_\a u^\a$ and ${\cal A}_\a\equiv h_\a{}^\b A_\b$.
In MHD the current $H_{\rm f}^\a$ is no longer conserved.  We propose to replace it by [notation as in Eqs.~\reff{w} and \reff{W}]
\be
\bar H_{\rm f}^\alpha = {}^*\!{W}^{\a\b}\, w_\b.
\label{Hf}
\ee
The proof that $\bar H_{\rm f}^\alpha$ is conserved is simple.  In view of Eq.~\reff{p2} we have
\be
\bar H_{\rm f}^\alpha{}_{;\a}=\sfr{1}{2}\,^*\!{W}^{\a\b}\, W_{\a\b},
\label{Hf1}
\ee
where  the antisymmetric part of $w_{\b;\a}$ is the only part that survives contraction with the antisymmetric $^*\!{W}^{\a\b}$.
Now $W_\a=0$  by Eq.~\reff{Wu=0},  so from Eq.~\reff{wdec} we see that \bea
W_{\a\b} &=& \varepsilon_{\a\b\g\d}\,u^\g V^\d,
\label{Wmag}
\\
^*\!{W}_{\a\b} &=& -u_\a V_\b + u_\b V_\a.
\label{Wstar}
\eea
It is obvious that the full contraction of indices between these two tensor vanishes identically; hence by Eq.~\reff{Hf1}  $\bar H_{\rm f}^\a{}_{;\a}=0$.

Two additional candidates for conserved helicity currents suggest themselves.  One is
$^*\!{F}^{\a\b}\, w_\b$; however, in view of Eqs.~\reff{Fstar}, \reff{etau} and~\reff{etaB}, this is no different from Moffat's fluid-magnetic helicity $H_{\rm fm}^\a$.  The second is
$ ^*\!{W}^{\a\b}\, A_\b$.
Writing $^*\! W_{\a\b}$ in terms of $w_\b$ and $^*\! F_{\a\b}$ in terms of $A_\b$ gives
\be
^*\!{W}^{\a\b}\, A_\b- {}^*\!{F}^{\a\b}\, w_\b=\sfr{1}{2}\big(\varepsilon^{\a\b\g\d}\, w_\d\, A_\b\big)_{;\g}
\ee
Since the r.h.s. is divergenceless (by antisymmetry of $\varepsilon^{\a\b\g\d}\, w_\d\, A_\b$), the  proposed helicity current is indeed conserved as a result of the conservation of $H_{\rm fm}^\a$.

It may also be seen that the conserved helicity (the zeroth component of $^*\!{W}^{\a\b}\, A_\b$ integrated over space) differs from the Moffat's fluid-magnetic helicity (the time component of $H_{\rm fm}^\a$ integrated over space)  by
\be
\sfr{1}{2}\int \big(\varepsilon^{\a\b\g\d}\, w_\d\, A_\b\big)_{;\g}
\,d\Sigma_\a = \sfr{1}{2}\oint \varepsilon^{\a\b\g\d}\, w_\d\, A_\b\,d\Sigma_{\a\g}
\label{int}
 \ee
where $d\Sigma_\a$ is a 3-volume element on the corresponding spacelike surface while $d\Sigma_{\a\g}$ is a 2-area element on the latter's boundary at infinity.
This looks gauge dependent.  However, assume we are able to isolate a ``physical'' part of $A_\b$.  We shall now show that the corresponding integral vanishes if the fluid extends to infinity.

It is convenient to think of the boundary at infinity as spherical.  Thus $d\Sigma_{\a\g}$ comprises only   temporal and radial components.  By the antisymmetry of $\varepsilon^{\a\b\g\d}$  only the angular components of $w_\d$ and $A_\b$ will contribute to the second integral in Eq.~\reff{int}.

Now asymptotically $\varepsilon_{\a\b\g\d}\sim r^2$; accordingly  the relevant components of $d\Sigma_{\a\g}$ vary as $r^2$ while $\varepsilon^{\a\b\g\d}\sim r^{-2}$. Thus the integral has the asymptotic behavior of $w_\d\, A_\b$ with $\b$ and $\d$ angular coordinates.  Now the phase $\varphi$ should depend on angular coordinates and not fall off with radius, so $w_\d$ does not decay with $r$.  However, $A_\b$ must fall off with increasing $r$.  Were this untrue, $F_{\a\b}$ with both indices angular variables would survive at large $r$ which would mean that the physical radial magnetic field falls off as $r^{-2}$  or slower.  But this would imply a nonzero magnetic monopole field which we may discard as unphysical.  Thus the integral in Eq.~\reff{int} vanishes.  It follows that the current $ ^*\!{W}^{\a\b}\, A_\b$ does \emph{not} furnish a new conserved helicity.

In conclusion, perfect MHD has three different conserved helicities corresponding to the currents defined in Eqs.~\reff{Hm}-\reff{Hf}.  The fluid helicity conservation law constructed on the basis of Oron circulation was first claimed in Ref.~\onlinecite{bekhelicity}, but only for stationary axisymmetric flow.  We have now lifted the symmetry restriction.    By contrast the claim in Ref.~\onlinecite{bekhelicity} that there is a pair of conserved fluid-magnetic helicities based on Oron's circulation has been here nullified by our demonstration that these are identical with the conserved Woltjer fluid-magnetic helicity.

\section{Bernoulli theorems}
\label{sec:bernoulli}

The Bernoulli theorem is well known from nonrelativistic perfect fluid mechanics; it actually appears in two types~\cite{LL}.  The type-1 theorem obtains in any stationary adiabatic flow: the sum of the specific enthalpy, specific bulk kinetic energy and gravitational potential is invariant \emph{along streamlines}.   The type-2 Bernoulli theorem states that for  isentropic stationary \emph{potential} flow, the above mentioned sum is a constant \emph{all over the flow}.  Relativistic formulations of both types are known~\cite{nov,carter}.  Less well appreciated is the existence of an additional Bernoulli theorem for \emph{each} space symmetry  shared by the flow and the gravitational potential (or geometry in the relativistic form).

Relativistic MHD also exhibits Bernoulli theorems.  For example, in any flow with both time and azimuthal Killing vectors, BEO exhibited a pair of type-1 Bernoulli theorems~\cite{BEO}.    It is quite clear from their work that similar theorems arise if the mentioned symmetries are replaced by some others,  but that two symmetries are needed together.  In the sequel we show, by exploiting the field description of this paper, that one can actually obtain one Bernoulli theorem of type-1\, and one of type-2 for each spacetime symmetry of the geometry and flow.

\subsection{Case with one spacetime symmetry}
\label{sec:onesymm}

\subsubsection{Type-1 Bernoulli theorems}
\label{sec:along}

Assume that the metric and the fluid 4-velocity possess a symmetry described by a single Killing vector $\xi^\alpha$.  This  $\xi^\a$ satisfies Killing's equation
\be
\nabla^\a\,\xi^\b + \nabla^\b\,\xi^\a = 0. \label{Killing}
\ee
Let us decompose the Killing vector as follows:
\be
\xi^\a = k\,u^\a + k^\a \qquad \bra{k = -\xi^\a\,u_\a\ ; \quad k^\a\,u_\a = 0}.\label{xidec}
\ee
Inserting Eq.~\reff{xidec} into Eq.~\reff{Killing} and contracting in turn with $u_\a\,u_\b$ and with $u_\a\,h_{\b}^{~\g}$ produces the independent equations
\bea
\dot k + k^\a\,\dot u_\a &=& 0\,, \label{k1}
\\
k\,\dot u^\a + \dot k^{\la \a \ra} - \D^\a k - k^\b\,\D^\a\,u_\b &=&0\,, \label{k3}
\eea
which we shall employ presently.

Further we record the Lie derivative of a scalar function $f$ and of the 4-velocity field $u^\a$ with respect to $\xi^\a$:
\bea
\mathscr{L}_\xi f &\equiv & \xi^\b\nabla_\b\, f=  k\,\dot f + k^\a\,\D_\a\,f ,
\label{kscalar}
\\
\mathscr{L}_\xi\, u^\a &\equiv& \xi^\b\nabla_\b\, u^\a-u^\b\nabla_\b\, \xi^\a 
\nonumber
\\
=&-&\left(\dot k + k^\b\,\dot u_\b\right) u^\a + k^\b\,\D_\b\,u^\a -\dot k^{\la \a \ra}\,.
 \label{lieu1}
\eea
Since the flow partakes in the symmetry $\xi^\a$ of the spacetime, $\mathscr{L}_{\xi}\,u^\a = 0$, which upon using the first Killing equation \reff{k1} in \reff{lieu1}, implies
\be \label{dotkcon}
\dot k^{\la \a \ra} = k^\b\,\D_\b\,u^\a.
\ee
Eliminating $\dot k^{\la \a \ra}$ between this equation and Eq.~\reff{k3} provides the convenient expression for the fluid's acceleration
\be
\dot u^\a = \D^\a\,\ln k + 2\,k^{-1}\,\omega^{\a\b}\,k_\b\,,
\label{accel}
\ee
which tells us that in the absence of vorticity the acceleration is the gradient of the ``potential'' $-\ln(\xi^\a u_\a)$.

Let us compare this result with Euler's equation~\reff{Euler3} with $W^\a=0$ and  Eq.~\reff{FB}:
\be
\dot u^\a =- \D^\a\,\ln \mu +(n\mu)^{-1}\epsilon^{\a\b\g}\,J_\b B_\g\,.
\ee
We infer that
\be
\bra{\mu\,n}^{-1}\epsilon^{\a\b\g}\,J_\b\,B_\g = \D^\a \ln\bra{k\,\mu} + 2\, k^{-1}\omega^{\a\b}\,k_\b\,.
\ee
Contracting this with $k_\b$ gives
\be
n^{-1}\epsilon^{\a\b\g}\,k_\a\,J_\b\,B_\g = \mu\,k_\a\, \D^\a \ln\bra{k\,\mu} = - {\cal D}\bra{k\,\mu} , \label{lhs}
\ee
where Eq.~\reff{kscalar} has been employed to achieve the second equality. The magnetic term here can be gotten from  Eq.~\reff{Hall} contracted with $k_\a$.  The expression involving  kinematical variables appearing therein can be deduced by contracting Eq.~\reff{Du}  with $k^\a \eta^\b$ and simplifying with use of Eq.~\reff{dotkcon}.
The result is
\bea
n^{-1}\epsilon^{\a\b\g}\,k_\a\,J_\b\,B_\g &=& -\zeta\big(\dot\eta^{\la \a \ra}\,k_\a+\eta^\a \dot k_{\la \a \ra}\big)
\nonumber
\\
&=& - {\cal D}\bra{\zeta\,\eta^\a\,k_\a} . \label{lhs2}
\eea
with the last equality following because both $k_\a$ and $\eta^\a$ are orthogonal to $u^\a$ so that $\dot\eta^{\la \a \ra}$ can be replaced by ${\cal D}\eta^{\alpha}$, etc.

Combining Eqs.~ \reff{lhs} and \reff{lhs2} we  arrive at the sought result:
\be
 {\cal D}\bra{\zeta\,\eta^\a\,k_\a - k\,\mu} = 0\, , \label{Bernoulli1}
\ee
which is the relativistic MHD Bernoulli theorem. In other words, we have shown that
\be \label{Bernoulli2}
\mu\bra{u_\a\,\xi^\a} + \bra{4\,\pi\,n}^{-1}\epsilon^{\a\b\g}\,\xi_\a\,b_\b\,B_\g = K\,,
\ee
where $K$ is constant along streamlines.  It seems not to be possible to prove that $K$ is a global constant in the absence of additional assumptions, so what we have found is a type-1 Benoulli theorem for each spacetime symmetry.

\subsubsection{Type-2 Bernoulli theorems}
\label{sec:type2}

Now choose coordinates in such a way that one of them, $x^\xi$, increases along the integral lines of $\xi^\a$.  As an analog of what is referred to as potential flow, we choose the following \emph{ansatz} for the phase in our formalism:
\be
\varphi=K x^\xi +H(x^R)\,.
\label{ansatz}
\ee
Here $H$ is some function of the coordinates other than $x^\xi$, which we denote collectively by $x^R$,  and $K$ is a constant.   We now show that this choice of $\varphi$ is consistent with the postulated symmetry.

 It is clear that $\varphi_{,\a}$ does not depend on $x^\xi$.
Thus because $u^\a$ shares in the symmetry, it follows from Eqs.~\reff{mu1} and \reff{nu=Cmu} that $\nu$ and $\mu$ are $x^\xi$ independent.  From Eq.~\reff{n1} we find, likewise, that $n/\sigma^2$ is $x^\xi$ independent.  We then gather from  Eqs.~\reff{rho1} and \reff{p1}  summed together that $\rho+p$ shares in this symmetry and then Eq.~\reff{mu} shows that $n$ and thus $\sigma^2$ do likewise.  Returning to Eqs.~\reff{rho1} and \reff{p1} we verify the symmetry of $\rho$ and $p$ separately.  All the thermodynamic quantities thus share in the symmetry. Finally from Eq.~\reff{eta1} follows that $\eta_\a$ is $x^\xi$ independent.    Eq.~\reff{4vel} then confirms that $u^\a$ comes out $x^\xi$ invariant as originally assumed.   The above argument does not prove that the ansatz~\reff{ansatz} is unique, but only that it does not conflict with expectations for the assumed symmetry. 

We now argue that the \emph{ansatz} \reff{ansatz} indeed represents potential flow.   Evidently $u^\a$ should  be a single valued vector; this requires single-valuedness of $H(x^R)$.  If in addition the coordinate $x^\xi$ is not compact, e.g. $\xi^\a$ is a time translation Killing vector, then $\varphi$ is single valued and the flow it represents does not contain circulation; it is potential flow.  Much the same can be said when $x^\xi$  is compact, as when $\xi^\a$ is an axial symmetry.  All loops around the axis of symmetry have then the same circulation,  and it is still possible, with care, to use a (multivalued) potential for this velocity field just as one may use a multivalued scalar potential to describe the magnetic field around an electric current filament.

Let us now substitute the ansatz ~\reff{ansatz} into the contraction of Eq.~\reff{4vel} with $\xi_\a$:
\be
\mu u_\a\,\xi^\a+\zeta\eta_\a\,\xi^\a=K.
\label{ber2}
\ee
Because $K$ here is a global constant, this is the promised type-2 Bernoulli theorem.  To recast it  into an alternative form more reminiscent of nonrelativistic versions of Bernoulli's theorem we substitute $u_\xi=u_\a\,\xi^\a$ from the above equation into $u^\a u_\a=-1$ to obtain
\bea
&\sfr12&\,g^{RS}u_R\,u_S+ \mu^{-1}(K-\zeta\eta_\a\,\xi^\a) g^{\xi R}u_R=
\nonumber
\\
-&\sfr12& \mu^{-2}g^{\xi\xi}(K-\zeta\eta_\a\,\xi^\a)^2-\sfr12\,.
\label{alter}
\eea
We observe from Eq.~\reff{eta} that  $\zeta\eta_\a\,\xi^\a$ is $\zeta$ independent as well as quadratic in the electromagnetic field [according to Eq.~\reff{beq}, $b^\a$ is linearly related to the magnetic field].

Let us pass to nonrelativistic MHD flow in a static, nearly flat spacetime. Focusing on the time translation symmetry, $\xi^\a=\delta_t^\a$,  we may write the squared space velocity as $v^2=g^{RS}u_R\,u_S$, $g^{\xi R}=g^{tR}\approx 0$,  $g^{\xi\xi}=g^{tt}\approx -(1-2\Phi_N)$ ($\Phi_N$ is the Newtonian potential), and $\rho=mn+e$ ($m$ is the rest mass per baryon and $e$ the internal energy density).  We see from Eq.~\reff{ber2} that in slow motion with negligible pressure and negligible magnetic effects, $K\approx -m$ so that we may define ${\cal E}$ by $K=-m(1+{\cal E})$, and assume $|{\cal E}|\ll 1$.  Similarly we should think of the nonrelativistic specific enthalpy $h\equiv (e+p)/mn$ and $\zeta\eta_\a\,\xi^\a/m$ as small compared to unity ($c=1$).  It then follows from Eq.~\reff{alter} to first order that
\be
\sfr12 v^2 +h+\Phi_N-(\zeta/m)\eta_\a\,\xi^\a={\cal E}\,.
\label{nonr}
\ee
This is the generalization of the classic nonrelativistic Bernoulli theorem to MHD. We see that ${\cal E}$ is the nonrelativistic energy per unit mass.

In order that the above results be really useful we must calculate $\eta_\a\,\xi^\a$, or equivalently $F_{\xi\beta}\, b^\b$.  But we have not succeeded in solving Eqs.~\reff{beq} and \reff{mj} for $b^\a$. This remains a problem for the future.   However, in a situation with two symmetries, $\xi_1^\a$ and $\xi_2^\a$, we can reach useful conclusions by comparison with BEO's type-1 theorem.  We now turn to this.

\subsection{Case with two spacetime symmetries}
\label{sec:twosym}

\subsubsection{Type-1 Bernoulli theorems}
\label{type-1}

We shall here recast BEO's results in a transparent form.  When the geometry and flow display two independent Killing vectors, $\xi_1^\a$ and $\xi_2^\a$, BEO find that
\bea
F_{\a\b}\,\xi_1^\a&=&A\,F_{\a\b}\,\xi_2^\a\,,
\label{FF}
\\
0 &=& F_{\a\b}\,\xi_1^\a\,\xi_2^\b \,,
\label{F03}
\\
B_\a&=&-Cn(u_\b\,\chi^\b u_\a+\chi_\a)\,,
\label{B1}
\\
B^2&=&C^2n^2[(u_\b\,\chi^\b)^2+\chi_\a\,\chi^\a]\,,
\label{B2}
\eea
where $A$ and $C$ are constants along streamlines, and
\be
\chi^\a\equiv \xi_1^\a - A\xi_2^\a\,.
\label{Xi}
\ee
(The dimensions of $A$ in the definition of $\chi^\a$ depend on the type of symmetries involved.)

BEO also find the  Bernoulli theorems ($i=1, 2$)
\be
\xi_i^\a\Big(\mu u_\a+ {B^2 u_\a+Cn\, u_\b \chi^\b\, B_\a\over  4\pi n }\Big)= K_i,
\label{K1}
\ee
where  $K_1$ and $K_2$ are again constants on each streamline.  BEO show that
\be
 \mu\,u_\a\,\chi^\a  = K_1-A K_2.
\label{D}
\ee
In all this BEO assume that the field component $F_{\a\b}\,\xi_1^\a\,\xi_2^\b $ vanishes asymptotically; Maxwell's equations then force it to vanish identically.  

Let us now compute $B_\b\,\xi_i^\b$ in Eq.~\reff{K1} by means of Eq.~\reff{B1}, and then eliminate $C^2 n^2$ in favor of $B^2$ by means of Eq.~\reff{B2}. After simplification and a cancellation we obtain the type-1 Bernoulli  theorems ($i=1, 2$)
\be
\mu\,u_\a\,\xi_i^\a+{ B^2\over 4\pi n}\left[u_\a \xi_i^\a\, \chi_\b\chi^\b-u_\a\chi^\a\,  \chi_\b \xi_i^\b \over (u_\a\chi^\a)^2+\chi_\a\chi^\a \right]=K_i\,.
\label{ber1}
\ee

\subsubsection{Type-2 Bernoulli theorems}
\label{Btype2}

Let us now imagine that the flow is described by the ansatz
\be
\varphi=K_1 x^{\xi_1} +K_1 x^{\xi_2} + H(x^R),
\label{newansatz}
\ee
where $K_1$ and $K_2$ are constants, $x^{\xi_1}$ and $x^{\xi_2}$  are the coordinates along integral lines of $\xi_1^\a$ and $\xi_2^\a$, respectively, and the $x^R$ are the two other coordinates.  By the same method leading to Eq.~\reff{ber2} we now get the two type-2 Bernoulli theorems ($i=1, 2$)
\be
\mu u_\a\,\xi_i^\a+\zeta\eta_\a\,\xi_i^\a=K_i\,. \label{ber3}
\ee
Obviously   $K_1$ and $K_2$ are also constant along streamlines, and so these two laws may well correspond to the two theorems, Eqs.~\reff{ber1}, which certainly remain valid  under the restricted flow described by Eq.~\reff{newansatz}.   (Note, however, that we cannot regard $A$ or $C$ as constant all over the flow.)   

To back up the above identification we immediately verify by use of the definition~\reff{eta}  that, regardless of the form of $b^\a$,
\be
\eta_\a\,\xi_1^\a = A\, \eta_\a\,\xi_2^\a\,.
\label{etaeta}
\ee
This is because
\be
\xi_1^\a\,F_{\a\b}\,b^\b = F_{\xi_1 R}\,b^R\,.
\ee
  A similar expression applies with $\xi_2^\a$, and so Eqs.~\reff{FF}-\reff{F03} give Eq.~\reff{etaeta} for any $b^\a$.    If the identification of Eq.~\reff{ber3} with BEO's result is justified, we must find the ratio of the magnetic term in theorem~\reff{ber1} with $i=1$ to that with $i=2$ to be exactly $A$, just as predicted by Eq.~\reff{ber3} in light of Eq.~\reff{etaeta}.  This is easily verified if cognizance is taken of Eq.~\reff{Xi}. Finally, if we substract $A$ times Eq.~\reff{ber3} as applicable with $\xi_2^\a$ from the same equation for $\xi_1^\a$,  Eq.~\reff{D} of BEO emerges.  Thus,  the identification with BEO is consistent, and the conserved expressions in Eq.~\reff{ber3} and \reff{ber1} must be the same.

Accordingly, we find in our formalism that
\be
\zeta\eta_\a\,\xi_i^\a={ B^2\over 4\pi n}\left[u_\a \xi_i^\a\, \chi_\b\chi^\b-u_\a\chi^\a\,  \chi_\b \xi_i^\b \over (u_\a\chi^\a)^2+\chi_\a\chi^\a \right],
\label{eta2}
\ee
which exhibits the expected quadratic character of $\eta_\a$ in the magnetic field. Putting this in Eq.~\reff{ber3} we get two type-2 Bernoulli theorems in explicit form.

\paragraph*{Example 1:}

Consider a situation with both a time Killing vector, $\xi_1^\a=\delta_t^\a$, and a translational spatial Killing vector in the $x$ direction, $\xi_2^\a=\delta_x^\a$.  We shall work nonrelativistically from Eq.~\reff{nonr} interpreted as a type-2 theorem, and assume that the magnetic energy per baryon, $B^2/8\pi n$, is not large compared to the kinetic energy per baryon $ v^2/2$. Then it is unnecessary to take into account terms of ${\cal O}(v^2)$ in the square brackets in Eq.~\reff{eta2} which would generate terms comparable to ${\cal O}(v^4)$, the like of which have already been neglected in Eq.~\reff{nonr}.  By the same token we drop in Eq.~\reff{eta2}  corrections to the Minkowski metric which are of ${\cal O}(\Phi_N)={\cal O}(v^2)$.  The resulting Bernoulli theorem to ${\cal O}(v^3)$ is
\be
\sfr12 v^2 +h+\Phi_N+{B^2\over 4\pi n m}\left({v_x+A  \over 2v_x+A }\right)={\cal E}\,.
\ee
We recall that in the last few equations $A$ is constant only along streamlines.

\paragraph*{Example 2:}

Still starting from Eq.~\reff{nonr} we shall replace the above $\xi_2^\alpha$ by an azimuthal Killing vector (cylindrical coordinates $\{r,z,\varphi\}$): $\xi_2^\a=\delta_\varphi^\a$. The streamline constant will here be denoted $\bar A$.   In the nonrelativistic limit we shall put $u_t\approx -1$ and $u_\varphi= r^2u^\varphi\approx r^2\Omega$, where $\Omega$ denotes the azimuthal angular velocity $d\varphi/dt$.  Again we neglect terms of ${\cal O}(\Omega^2)$ in the square brackets in Eq.~\reff{eta2}, so obtaining
\be
\sfr12 v^2 +h+\Phi_N+{B^2\over 4\pi n m}\left({\Omega+\bar A  \over 2\Omega+\bar A }\right)={\cal E}\,.
\ee

\paragraph*{Example 3:}

We return to the situation with time and $x$ symmetries.  In Eq.~\reff{ber3} with $\xi_i^\a=\xi_2^\a=\delta_x^\a$ we shall use the expression~\reff{eta2}.  Nonrelativistically we approximate $u_\a\xi_1^\a\approx-1$ and $u_\a\xi_2^\a\approx v_x$ where $v_x$ is the component of the ordinary velocity in the symmetry direction.  We take the metric everywhere as Minkowski's and neglect $h$ in comparison with unity.  The result is
\be
v_x-{B^2\over 4\pi n m}\left({v_x+A  \over A(2v_x+A) }\right)={\cal P}\,,
\ee
where ${\cal P}\equiv K_2/m$ is the linear momentum per unit mass.

\subsubsection{New conserved circulation}
\label{C}
 
Eq.~\reff{eta2} gives the projections of $\zeta\eta_\a$ onto the two Killing vectors.
Can we reconstruct $\zeta\eta_\a$ fully from this ?  For this purpose we could try finding the projections of   $\eta_\a$ onto two independent vectors, both orthogonal to the plane spanned by $\xi_1^\a$ and $\xi_2^\a$.  We use the following two:
\bea
Q_1^\a&\equiv& \Xi_{\g\d}\,  \Xi^{\g\d} u^\a -2 u^\b \Xi_{\b\g} \Xi^{\a\g} 
\\
Q_2^\a&\equiv&\varepsilon^{\a\b\g\d}u_\b\, \Xi_{\g\d}=-\epsilon^{\a\g\d}\Xi_{\g\d}
\label{q2}
\\
\Xi^{\a\b}&\equiv& \xi_1^\a\, \xi_2^\b - \xi_2^\a\, \xi_1^\b
\eea
It is immediately verified that $Q_1^\a\,\xi_{i\a}=Q_2^\a\,\xi_{i\a}=0$ for $i=1,2$. Now since $Q_1^\a$ is a linear combination of $u^\a,\xi_1^\a$ and $\xi_2^\a$, it follows immediately from the full antisymmetry of the Levi-Civitta tensor that $Q_1^\a$ and $Q_1^\a$ are orthogonal and hence independent as required.  Therefore, it is no curtailment of generality to write
\be
\zeta\eta^\a={ B^2\over 4\pi n}\left[u^\a \, \chi_\b\chi^\b-u^\b\chi_\b\,  \chi^\a  \over (u_\g\chi^\g)^2+\chi_\g\chi^\g \right]+a_1Q_1^\a+a_2Q_2^\a,
\label{neweta}
\ee
since one recovers from this ansatz both projections~\reff{eta2}, and it has enough freedom left  to represent \emph{any} vector in the space complementary to the symmetry directions.   Both scalars $a_1$ and $a_2$ must depend quadratically on $B^\a$ because $\eta^\a$ is quadratic in it.  In fact, by demanding $\eta_\a u^\a=0$ and noticing that $Q_2^\a u_\a=0$ we find that 
\be
a_1=-{B^2/(4\pi n)\over  \Xi_{\a\b}\,  \Xi^{\a\b} +2(u_\a\Xi^{\a\b} )^2}\ .
\ee

We shall now argue that $a_2$ must vanish on grounds of parity.   By Eq.~\reff{eta1} the space part of $\eta^\a$  must be a true 3-vector (changing sign under a space inversion) since $\varphi$ should be a true scalar---invariant under inversion.   Both $u^\a$'s space part and those of spatial Killing vectors must also be true 3-vectors.  And by Eq.~\reff{FF}, $A$ must be a true scalar.  Hence any scalar product like $u_\a \xi_i^\a$,  $u_\a \chi^\a$, $\xi_{i\a}\,\xi_j^\a$ or $\chi_\a\,\chi^\a$ must be a true scalar.     Thus the square brackets  in Eq.~\reff{neweta} enclose a 4-vector whose space part is a true 3-vctor.   Likewise, the space part of $Q_1^\a$ is a true 3-vector, while $a_1$ is a true scalar.  It follows that  $a_2Q_2^\a$ must be a true 3-vector.  However, it is clear from the second form in Eq.~\reff{q2} that the space part of $Q_2^\a$ is actually a pseudo 3-vector, so that $a_2$ should be a pseudoscalar.  Yet no such can be built out of $B^\a$ (whose space part is a pseudo 3-vector) and the $\xi_i^\a$ which is also quadratic in the $B$ field.  The closest we come is $B_\a\xi_i^\a\cdot (\epsilon_{\b\g\d} B^\b \xi_1^\g \xi_2^\d)$ with $i=1$ or $i=2$.  However, the factor in parenthesis here  is just $F_{\a\b}\,\xi_1^\a\xi_2^\b$, which vanishes by Eq.~\reff{F03}.  We thus conclude that the $Q_2^\a$ term in $\zeta\eta^\a$ must be absent, and that we have reconstructed the full $\eta^\a$ field.

The above does not inform us further on Bernoulli theorems because $Q_1^\a$ is orthogonal to both Killing vectors.   But somewhat surprisingly it gives a further circulation conservation law.  According to Eq.~\reff{Gamma}, the circulation of $\mu u_\a+\zeta \eta_\a$ is conserved, and according to BEO the circulation of $\mu u_\a$ plus just the first term on the r.h.s. of Eq.~\reff{neweta} is separately conserved.  We conclude that
\be
\Gamma_{B^2} = \oint a_1\, Q_{1\a} dx^\a
\ee
is a new conserved circulation for MHD flows with two spacetime symmetries.

\paragraph*{Example:}

Consider again the situation with both a time Killing vector $\xi_1^\a=\delta_t^\a$ and a spatial one in the $x$ direction $\xi_2^\a=\delta_x^\a$.  In  flat spacetime but working fully relativistically we have $\Xi_{\g\d}\,  \Xi^{\g\d}=-2$ and $u^\b \Xi_{\b\g} \Xi^{\a\g} = u_t \,\xi_1^\a-u_x\,\xi_2^\a$.  Thus $Q_1^\a=-2u^\a-2u_t\xi_1^\a+2u_x\xi_2^\a$ so that $Q_{1\a}dx^\a=-2(u_y dy +u_z dz)$ where $y$ and $z$ are the usual Cartesian coordinates orthogonal to the symmetry direction.  From the normalization $u_\a\,u^\a=-1$ it follows that $(u_x \,\xi_1^\a-u_t\,\xi_2^\a)^2=u_t^2-u_x^2=1+u_y^2+u_z^2$.  With all these pieces it follows that the circulation
\be
\oint {B^2\over 4\pi n}{u_y dy +u_z dz\over u_y^2+u_z^2}
\ee
around an arbitrary loop moving with the flow is conserved.

\section{Summary}

Perfect fluid flow can be represented as the long wavelength behavior of scalar field dynamics. When the fluid is charged, it is represented by a complex scalar field minimally coupled to the electromagnetic gauge potential.  In this paper we have provided, in the framework of general relativity, a complex scalar field representation for the flow of highly conducting but neutral magnetized perfect fluid (perfect magnetohydrodynamics).   The coupling to electromagnetism is via a vector field distinct from the electromagnetic gauge vector, but the theory has a full $U(1)\times U(1)$ gauge symmetry. The scalar field's  self-interaction determines the fluid's equations of state. 

 The principal advantage of this theory  is that it leads directly to Oron's conserved MHD circulation which plays the role analogous to that of Kelvin's circulation in pure fluid flow.    Here we obtain the Oron circulation without the need to assume two spacetime symmetries.  Additionally, with two symmetries present we find an entirely new circulation involving the magnetic energy per particle and the velocity field.  We have discussed the structure of the conserved helicity current which is associated with Oron's circulation.  Finally we have put in evidence the existence of a pair of Bernoulli-like theorems which are associated with each type of spacetime symmetry exhibited by the MHD flow.  In the case of two simultaneous symmetries we are able, by comparing with old results, to obtain explicit forms for the Bernoulli conserved quantities.

\acknowledgments

We thank Marcelo Schiffer for a suggestion.   GB thanks the Hebrew University for support.  This research was supported by grant  694/04  of the Israel Science Foundation, established by the Israel Academy of Sciences and Humanities.

\end{document}